\documentclass[12pt,reqno]{amsart}
\usepackage{amsmath,amssymb,amsfonts,amsthm}
\usepackage[mathscr]{eucal}
\usepackage[all]{xy}
\usepackage{hyperref}
\usepackage{setspace}
\allowdisplaybreaks

\author{I.Yu.~Karataeva, S.L.~Lyakhovich, I.A.~Retuntsev}

\address{Physics Faculty, Tomsk State University, Lenin ave. 36, Tomsk 634050, Russia}

\email{karin@phys.tsu.ru, \, sll@phys.tsu.ru, \, retuntsev.i@phys.tsu.ru}

\title{Topological string as massive spinning particle \\ in three dimensions}
\begin{document}
\maketitle

\begin{abstract}
A model is proposed for a classical bosonic string in $d=3$ Minkowski space with an action functional that includes Gauss and mean world-sheet curvature.
The Lagrangian is invariant under $3d$  Poincar\'e  transformations modulo total divergence.
In addition to the diffeomorphism, the action enjoys the extra gauge symmetry with the second derivatives of the scalar gauge parameter.
This symmetry gauges out all the local degrees of freedom (DoF's), while some global DoF's survive. The Hamiltonian constrained analysis confirms that the model does not have any local DoF.  The world sheet of the string turns out to be a cylinder with time-like axis. The global DoF's of this string describe one single irreducible massive $3d$ particle with spin. The particle momentum is a conserved vector directed along the cylinder axis while the momentum square is a fixed constant determined by the parameters in the action. The total angular momentum is a conserved vector that defines the position of the axis of the cylinder whose specific value is defined by initial data, while the spin, being the product of momentum and angular momentum is fixed by the parameters in the string Lagrangian.

\end{abstract}

\label{sec:intro}
\section{Introduction}
The known models of classical strings typically describe an infinite spectrum of particles with various spins and masses.
The degrees of freedom (DoF's) of these particles come from the oscillations of the string.
From this perspective, classical spinning particle could be thought of as a specific truncation of the string model such that fixes, and/or gauges out, infinite number of the string oscillating modes. The remaining finite number of evolving modes should describe one single irreducible classical particle with spin.

In this article, we propose the modification of the Nambu-Goto action in $d=3$ by the terms that fix Gauss and mean curvature of the world sheet. Inclusion of these terms leads to the additional gauge symmetry of the action that gauges out all the local DoF's of the string. The world sheet of this string is a right circular cylinder with the time-like axis. The cylinder radius is fixed by the constant included in the action, while the position and direction of the axis is defined by initial data. Such a rigid world sheet obviously does not oscillate, so the model is a topological $2d$ field theory having no local DoF's. The global DoF's of the model are related to the position  of the cylinder in $3d$ Minkowski space.  These data turn out to describe classical dynamics of irreducible $d=3$ massive spinning particle.

The connection between right circular cylindric world-sheet and world-line of the irreducible classical massive spinning particle is established in the article \cite{WS2017} in various dimensions, including $d=3$. To make this article self-contained, below, we briefly explain how the irreducible classical spinning particle dynamics are described by this world sheet.

At quantum level, $3d$ spinning particle is characterised by the two conditions:

\noindent
First, the irreducible representation of Poincar\'e group acts on the space of quantum states of the particle.
This means that the Casimir operators of the group are multiples of unit
\begin{equation}\label{Poincare}
  [\hat{P}_\mu,\hat{P}_\nu] =0\,,\quad[\hat{J}_\mu,\hat{J}_\nu]=i\hbar\epsilon_{\mu\nu\lambda}\hat{J}^\lambda\,, \quad [\hat{P}_\mu, \hat{J}_\nu]=i\hbar\epsilon_{\mu\nu\lambda}\hat{P}^\lambda \, ;
\end{equation}
\begin{equation}\label{Casimir}
  \hat{P}^2=-m^2\, , \quad (\hat{P}, \hat{J}) = ms \, .
\end{equation}
Second, the position of the particle in Minkowski space is represented by the operators $\hat{x}^\mu$ acting on this space of states. The coordinates $\hat{x}^\mu$ are transformed by the representation of the group, and they commute to each other because the position of the particle is supposed measurable
\begin{equation}\label{x}
  [\hat{x}^\mu, \hat{x}^\nu]=0\, , \quad [\hat{x}^\mu,\hat{P}_\nu]=i\hbar\delta^\mu_\nu\, , \quad [\hat{x}^\mu, \hat{J}^\nu]=i\hbar\epsilon^{\mu\nu\lambda}\hat{x}_\lambda \, .
\end{equation}
Once the coordinate and momentum operators act on the space of states, they define orbital momentum of the particle
\begin{equation}\label{M}
  \hat{M}^\mu=\epsilon^{\mu\nu\lambda}\hat{x}_\nu \hat{P}_\lambda \, , \quad  [\hat{x}^\mu, \hat{M}^\nu]=i\hbar\epsilon^{\mu\nu\lambda}\hat{x}_\lambda \, .
\end{equation}
Momenta $\hat{M}$ and $\hat{P}$  by construction generate Poincar\'e algebra,
\begin{equation}\label{PoincareM}
  [\hat{M}_\mu,\hat{M}_\nu]=i\hbar\epsilon_{\mu\nu\lambda}\hat{M}_\lambda\,, \quad [\hat{P}_\mu, \hat{M}_\nu]=i\hbar\epsilon_{\mu\nu\lambda}\hat{P}^\lambda \, .
\end{equation}
Let us introduce the spin operator as the difference between the total momentum $\hat{J}$ and orbital momentum $\hat{M}$. By construction, spin operators commute with $\hat{x}$ and $\hat{P}$, and they commute to each other as Lorentz generators,
\begin{equation}\label{Spin-oper}
  \hat{S}^\mu=\hat{J}^\mu-\hat{M}^\mu\, , \quad [\hat{S}^\mu,\hat{P}^\nu]=[\hat{S}^\mu,\hat{x}^\nu]=0,\quad [\hat{S}^\mu,\hat{S}^\nu]=i\hbar\epsilon^{\mu\nu\lambda} \hat{S}_\lambda\,.
\end{equation}
Given the above definition of spin operator, $\hat{S}^2$ commutes to all the generators $\hat{P}, \hat{J}$ of the irreducible representation. Hence, it is a multiple of unit,
\begin{equation}\label{Casimir_S}
 \hat{S}^2=\text{sign}(\varrho)\varrho^2\, ,
 \end{equation}
 where $\rho$ is a real constant.
This relation is consistent with the conditions (\ref{Casimir}) only if
\begin{equation}\label{intr_2}
s^2 + \text{sign}(\varrho)\varrho^2\geq0 .
 \end{equation}

Substituting the spin operator $\hat{S}^\mu$ (\ref{Spin-oper}) into irreducibility condition (\ref{Casimir_S}), we arrive at the relation  between Poncar\'e generators $\hat{P}, \hat{J}$ and particle position $\hat{x}$
\begin{equation}\label{WS_PJ}
(\hat{J}-\hat{M})^2=\hat{J}^2-2\epsilon^{\mu\nu\lambda}\hat{J}_\mu \hat{x}_\nu \hat{P}_\lambda+\hat{x}_\mu \hat{P}_\nu \hat{x}^\nu \hat{P}^\mu-\hat{x}_\mu \hat{P}_\nu \hat{x}^\mu \hat{P}^\nu=\text{sign}(\varrho)\varrho^2\,.
 \end{equation}
The identical relations (\ref{Casimir}) and  (\ref{WS_PJ}) provide irreducibility of Poincar\'e group representation in the space of states of the particle having position in Minkowski space.
Let us discuss now consequences of the irreducibility conditions (\ref{Casimir}),  (\ref{WS_PJ}) at classical level.
The operators of total momentum and angular momentum  are replaced by the corresponding classical conserved quantities $P, J$, while the operators of particle position turns into the coordinates of Minkowski space point $x$. Then, relations (\ref{Casimir}) reduce  to constraints on classical momentum and total angular momentum
\begin{equation}\label{Casimir_class}
  P^2=-m^2\, , \quad (P, J)= ms \, .
\end{equation}
Given the classical conserved quantities $P,J$, relation (\ref{WS_PJ}) reduces at classical level (with account of (\ref{Casimir_class})) to the equation that restricts position $x$ of the particle
\begin{equation}\label{WS_PJ_2}
J^2+(x,P)^2+m^2x^2-2 (x,[P,J])=\text{sign}(\varrho)\varrho^2\,.
\end{equation}
Here and below we use the square brackets of the classical vectors as abbreviation of vector product,
\begin{equation}\label{VectProdDef}
  [a,b]^\mu=\epsilon^{\mu\nu\lambda}a_\nu b_\lambda\,\quad \epsilon^{012}=-1\,.
\end{equation}
Relation (\ref{WS_PJ_2}), being quadratic form of particle's coordinates, defines a surface in Minkowski space. The conserved quantities $P$ and $J$ are involved in the equation of surface as constant parameters defined by initial data. The position  of irreducible classical massive spinning particle has to belong to this surface.

Let us demonstrate that algebraic equation (\ref{WS_PJ_2}) defines the right circular cylinder with time-like axis. It is convenient to introduce instead of $P$ and $J$ two other constant vectors $n$ and $y$,
\begin{equation}\label{PJ_ny}
y^\mu=\frac{1}{m^2}[P, J]^\mu,\quad n^\mu=\frac{1}{m}P^\mu \quad\Leftrightarrow\quad J^\mu=m[y, n]^\mu-sn^\mu,\quad P^\mu=mn^\mu.
 \end{equation}
Once $P$ and $J$ are constrained by relations (\ref{Casimir_class}), the vectors $n$, $y$ are orthogonal to each other and $n$ is normalized,
\begin{equation}\label{PJ_ny_2}
n^2=-1,\quad (n, y)=0\,.
 \end{equation}
In terms of normalized constant vectors $n$ and $y$, relation (\ref{WS_PJ}) reads
 \begin{equation}\label{WS_ny}
(x-y)^2+(n,x)^2=r^2,\quad r=\frac{1}{m}\sqrt{s^2 + \text{sign}(\varrho)\varrho^2}
 \end{equation}
It is the canonical form of equation that defines a circular cylinder of radius $r$ in 3d Minkowski space. The time-like unit vector $n$ is directed
along the axis of cylinder. The vector $y$, being orthogonal to $n$, connects the origin of reference system with the axis of cylinder. Once $n$ is time-like, $y$ is space-like. By Eq. (\ref{WS_ny}), the vector $n = p/m$ defines the direction of the cylinder axis, while the vector $y$ specifies the position of the axis in space. A cylinder of fixed radius with any position of the time-like axis is admissible. In this way, $n$ and $y$ parameterize the variety of all possible cylinders with time-like axis and fixed radius.
\par
The case of $s =\varrho=0$ obviously corresponds to a spinless point particle, so the paths cannot be anything but the straight lines. With non-vanishing spin, $s\neq 0$, the particle paths should be cylindric lines. The latter fact is sufficient to completely define the classical dynamics, including the equations of motion for the spinning particle in d=3 \cite{WS2017}.

Since the world line of an irreducible particle $x(\tau)$ with given conserved quantities belongs to the surface (\ref{PJ_ny}), we can repeatedly take time derivatives from the equation (\ref{PJ_ny}) and eliminate the constants $ n,y$ from these differential consequences, expressing them as terms of $\dot{x},\ddot{x}, \dddot{x}, \dots$. This leads to higher order differential equations for the world line $x(\tau)$ \cite{WS2017}. All the general world-lines on the same cylinder are gauge equivalent by construction, so in fact the classical dynamics of the spinning particle is rather a world-sheet than a world-line. These higher derivative equations are not Lagrangian. Introducing extra variables to depress the order, these equations can be reduced \cite{WS2017} to the known Lagrangian models of irreducible spinning particle in $d=3$ \cite{Anyon1996}, and $d=4$ \cite{UM1996}. The quotient of the Hamiltonian constrained surface of this particle model by gauge orbits coincides with the corresponding co-adjoint orbit of the Poincaré group \cite{UM1996}.

Since the dynamics of the single irreducible classical spinning particle is described by the specific world sheet,
it seems natural to find the action functional for corresponding string. So, we switch from the world-line $x(\tau)$ to the world-sheet $x(\tau,\sigma)$ and seek for the action such that has the circular cylinder with time-like axis as the extreme surface.

 In Euclidean space, the right circular cylinder is defined \cite{Geom_1} by the differential equations
 \begin{equation}\label{Gauss}
   K=0 \, ,
 \end{equation}
 \begin{equation}\label{Mean-r}
   2H+\frac{1}{r} =0\, , \quad r=const,
 \end{equation}
where $K$ and $H$ is the Gauss curvature and mean curvature of the surface, respectively. The first condition means that one of principal curvatures vanishes of the world sheet, so it is a ruled surface \cite{Geom_3}. Condition (\ref{Mean-r})  means that non-vanishing principal curvature is a constant everywhere on the world sheet. In this work we are interested in time-like surfaces in Minkowski space, so equations (\ref{Gauss}) define ruled surface for the same reasons as in Euclidean case.  The rules can be time-like, space-like, or null. The first case corresponds to the right circular cylinder with time-like axis, while the second one defines the hyperbolic cylinder with the space-like axis \cite{Geom_2} . The third case can occur only if the mean curvature vanishes, for example it can be a parabolic cylinder with the null axis \cite{Walrave}. The latter case is interesting because such a surface is the world sheet of a continuous helicity spinning particle in $3d$ Minkowski space \cite{WS2022_1}. Assuming Minkowski signature of the world sheet, and imposing corresponding periodicity conditions on $x$ as discussed below, we can stick with the right circular cylinder.

The first goal of the paper is to find an action functional for a string in $3d$ such that the world sheet equations reduce to (\ref{Gauss}), (\ref{Mean-r}). Construction of the action is considered in the next section. One can fix the mean curvature (\ref{Mean-r}) without introducing the auxiliary fields, while the Lagrange multipliers are  needed to enforce both equations (\ref{Gauss}), (\ref{Mean-r}) simultaneously. The key issue is to ensure that these additional fields are gauged out, otherwise the model will not correspond to an irreducible particle, even if the world sheet obeys the equations (\ref{Gauss}), (\ref{Mean-r}). The action turns out having the additional gauge symmetry, besides the diffeomorphism. Due to this symmetry, the Lagrange multipliers do not bring any local degree of freedom to the theory, and the gauge invariant dynamics reduces to the equations of the world sheet (\ref{Gauss}), (\ref{Mean-r}). Also notice that the equations (\ref{Gauss}), (\ref{Mean-r}) are not independent, there is Noether identity between them.  The identity is deduced in the Appendix. In Section 3, we work out Hamiltonian constrained formalism for this string, and see that it is a topological theory indeed, having no local degree of freedom.
 The global degrees of freedom, being the parameters of the cylinder, describe the dynamics of irreducible spinning particle as we explain above in the introduction.
\hspace{0.2cm}

\noindent
\textbf{Notation and agreements.}
The linear coordinates on Minkowski space are denoted $x^\mu, \, \mu=0,1,2$. Minkowski metric is mostly positive
\begin{equation}\label{eta}
\eta_{\mu\nu}=\text{diag}(-1,1,1).
\end{equation}
Local coordinates on the world sheet are denoted $\tau,\sigma$, or collectively $\xi^a, \, a=0,1;\, \xi^0=\tau, \xi^1=\sigma $.
The coordinate $\sigma\in[0,2\pi]$ is of the angle-type, and the world sheet is periodic in $\sigma$, $x(\tau,\sigma)=x(\tau,\sigma+2\pi)$.
Integration over the world-sheet is assumed within finite time interval $\tau\in[\tau_1,\tau_2]$ while $\sigma\in[0,2\pi]$,
\begin{equation}\label{int}
  \int d^2\xi \, f(\xi)=\int_{\tau_1}^{\tau_2}d\tau\int_0^{2\pi}d\sigma f(\tau,\sigma) \, .
\end{equation}
We adopt the usual notation for tangent vectors to the world sheet, $\partial_0x=\dot{x},\, \partial_1 x=\acute{x}$.
The induced metric
\begin{equation}\label{h-ab}\begin{array}
{c}\displaystyle
h_{ab}=(\partial_a x, \partial_b x)=\eta_{\mu\nu}\partial_a x^\mu \partial_b x^\nu\,,
\end{array}\end{equation}
is supposed to have signature $(-+)$, and $\dot{x}{}^2<0,\, \acute{x}{}^2>0 $, hence $h=\det{h_{ab}}<0$.
Given the periodic conditions, the space-like rules are inadmissible for the world sheet  with zero Gauss curvature (\ref{Gauss}).

The unit world-sheet normal $n$ is defined as the normalized vector product of two tangent vectors,
\begin{equation}\label{n}
\mathfrak{n}=-\frac{[\dot{x}, \acute{x}] }{\sqrt{-h}}\,, \quad
\mathfrak{n}^\mu
=\frac{1}{2\sqrt{-h}}\varepsilon^{ab}\varepsilon^{\mu\nu\rho}\partial_a x_\nu \partial_b x_\rho\,,\quad \varepsilon^{01}=\varepsilon^{012}=-1\,, \quad
\mathfrak{n}^2=1.
\end{equation}
The second fundamental form of the world-sheet is defined as the projection of second derivative of the string position to the normal vector,
\begin{equation}\label{b}\begin{array}
{c}\displaystyle
\mathfrak{b}_{ab}=( \partial_a\partial_b x,\mathfrak{n})=\eta_{\mu\nu} \partial_a\partial_b x^\mu \mathfrak{n}^\nu \, .
\end{array}\end{equation}
We adopt the following definitions for the world-sheet Gauss curvature $K$ and mean curvature $H$
\begin{equation}\label{KH-def}\begin{array}
{c}\displaystyle
K=\frac{\mathfrak{b}}{h}\,,\qquad 2H=h^{ab}\mathfrak{b}_{ab}\,, \quad \mathfrak{b}=\det{\mathfrak{b}_{ab}} \, .
\end{array}\end{equation}
The discriminant of the equation for principal curvatures of the world sheet is denoted
\begin{equation}\label{D-def}
\mathcal{D}=H^2-K\,, \quad \mathcal{D}\geq 0\, .
\end{equation}
 The discriminant is strictly positive if the principal curvatures are different.  If the principal curvatures coincide, $\mathcal{D}=0$. This special case is irrelevant to the subject of this article.

We denote $\nabla_a$ the covariant derivative that respects $h_{ab}, \nabla_a h_{bc}=0$.

Gauss and Peterson--Mainardi--Codazzi equations read\footnote{
Curvature tensor is defined as
$
R^a{}_{b\,cd} =
\partial_c \Gamma^a_{bd}
-\partial_d \Gamma^a_{bc}
+\Gamma^a_{cs}\Gamma^s_{db}
-\Gamma^a_{ds}\Gamma^s_{cb},
$ where $\Gamma$ is the Christoffel symbols for the metric $h_{ab}$.
}
\begin{equation}\label{GPMCe}
R_{ab\,cd}= K (h_{ac}h_{bd}-h_{ad}h_{bc}),
\qquad
\varepsilon^{ab}\nabla_a\mathfrak{b}_{bc}=0.
\end{equation}
We use the tensor $\mathfrak{B}^{ab}$ which is dual, in a sense, to the second fundamental form $\mathfrak{b}_{cd}$,
\begin{equation}\label{B}
\mathfrak{B}^{ab}=h^{-1}\varepsilon^{ac}\varepsilon^{bd}\mathfrak{b}_{cd} \, ,\quad \mathfrak{B}^{ac}\mathfrak{b}_{cb}=K \delta^{a}_{\,\,\,b}\, , \quad h_{ab}\mathfrak{B}^{ab}=2H\,.
\end{equation}
Tensor $\mathfrak{B}^{ab}$ is transverse due to Peterson--Mainardi--Codazzi equations (\ref{GPMCe})
for $\mathfrak{b}_{ab}$,
\begin{equation}\label{divB}
  \nabla_a \mathfrak{B}^{ab}=0\,.
\end{equation}
Two scalar second order differential operators can be constructed, given the transverse second rank tensors $h^{ab}, \, \mathfrak{B}^{ab}$,
\begin{equation}\label{LAP-def}
\Delta=h^{ab}\nabla_a\nabla_b = \nabla_a h^{ab}\nabla_b \,,\quad \widetilde{\Delta}=\mathfrak{B}^{ab}\nabla_a\nabla_b = \nabla_a \mathfrak{B}^{ab}\nabla_b\,.
\end{equation}
Both of these Laplace operators are involved in the gauge symmetry transformations constructed in the next Section.

\section{Action, equations, and gauge symmetry}
\label{sec:2}
In this section we  construct the action of the string in $d=3$ Minkowski space such that the extreme world sheet is defined by the equations (\ref{Gauss}), (\ref{Mean-r}). Given the action, we study gauge symmetry, and count degrees of freedom.

We begin with Nambu-Goto action. Due to the diffeomorphism, there are two Noether identities among three Lagrangian equations. The only remaining independent equation is the condition of zero mean curvature \cite{BH}, $H=0$. To have a cylinder as a world sheet rather than a minimal surface, we need a non-zero constant $H$ (\ref{Mean-r}). The equation (\ref{Mean-r}) appears if one more term is added to the Nambu-Goto action
\begin{equation}\label{S-WZ}\begin{array}
{c}\displaystyle
   S_1=-
\int d^2\xi
   \sqrt{-h}\bigg[\,\frac{1}{2\pi\alpha'}+\frac{\gamma}{3}(x,\mathfrak{n})\,\bigg]\,,
\end{array}\end{equation}
where $\alpha', \gamma$ are constant parameters of the model, and $\mathfrak{n}$ is a normal to the world sheet (\ref{n}).
 The added term is invariant under diffeomorphism, and Lorentz transformations of the target space. Poincar\'e translation changes this term by a total divergence.

Because of reparameterisation invariance, Lagrangian equations obey two Noether identities
\begin{equation}\label{GI-S-WZ}
  \partial_a x^\mu\frac{\delta S_1}{\delta x^\mu}\equiv 0\, ,\quad a=0,1\,.
\end{equation}
Since the projection of the equations identically vanishes on the tangential direction to the world sheet, the only nontrivial equation is the normal projection,
\begin{equation}\label{NormalS1}
  \mathfrak{n}^\mu\frac{\delta S_1}{\delta x^\mu}\equiv \frac{1}{\pi\alpha'}\sqrt{-h}\bigg(H-\pi\alpha'\gamma\bigg)= 0\, .
\end{equation}
As one can see, the extreme surface for the string action (\ref{S-WZ}) is the constant mean curvature  world sheet (\ref{Mean-r}).
Obviously, the second order equation describes single local DoF by Lagrangian count, much like the Nambu-Goto string in $d=3$.
So, the last term in the string action (\ref{S-WZ}) can be understood as Poincar\'e invariant constant background force directed by the normal to the string world sheet. The Euclidean counterpart has a different interpretation: it is the energy of the soap film in the field of constant gravity, see \cite{Action_2}. From the perspective of geometry, the constant mean curvature surfaces in $3d$ Minkowski is well studied subject, see \cite{Geom_2}, \cite{Walrave}, \cite{Lopez} for review and further references, though the variational principle (\ref{S-WZ}) has been previously unknown for Minkowski case, to the best of our knowledge.

The term $\sqrt{-h}K$ is a total divergence, so it is inessential for the action. Inclusion of the mean curvature $\sqrt{-h}H$ into the action (\ref{S-WZ}) adds the Gauss curvature to the field equation (\ref{NormalS1}),
\begin{equation}\label{S-WZ2}\begin{array}
{c}\displaystyle
   S_2=-
\int d^2\xi
   \sqrt{-h}\bigg[\,\frac{1}{2\pi\alpha'}+ \beta H + \frac{\gamma}{3}(x,\mathfrak{n})\,\bigg]\,,
\end{array}\end{equation}
\begin{equation}\label{NormalS2}
  \frac{\delta S_2}{\delta x^\mu}\equiv \frac{1}{\pi\alpha'}\sqrt{-h}\bigg(H +\pi\alpha'\beta K - \pi\alpha'\gamma\bigg)\mathfrak{n}_\mu = 0\, .
\end{equation}
So, by the modifications (\ref{S-WZ2}) of the Nambu-Goto action, we can  fix on shell at most the linear combination of the mean and Gauss curvature, not each of the curvatures.

Also notice, that discriminant $\mathcal{D}$ (\ref{D-def}) is the scalar with the second order derivatives of the world sheet such that $\sqrt{-h}\mathcal{D}$ does not reduce to a total derivative, unlike Gauss curvature $\sqrt{-h}K$. The remaining term $\sqrt{-h}H^2$ is well-known in geometry and describes so-called Willmore surfaces \cite{WSurf_Eucl}, \cite{WSurf_Mink}. A.M.~Polyakov proposed \cite{RS1} to add this term to the Nambu-Goto action to account for the fine structure of the strings at quantum level. Inclusion of this term also has consequences for the classical Regge trajectories \cite{RS2}. Inclusion of this term into the action (\ref{S-WZ2}) does not fix Gauss curvature on shell. As a result, this term would not lead to the string having a single particle in the classical spectrum.

The most obvious way to complement the equation of constant mean curvature (\ref{Mean-r}) with the condition of zero Gaussian curvature (\ref{Gauss}) is to include $K$ with the Lagrange multiplier  in the action (\ref{S-WZ}). This doesn't work: we get the equation $K=0$, but the mean curvature equation gets a right-hand side that includes the Lagrange multiplier at $K$. As a result, the world sheet will not be a surface of constant mean curvature for this action. The same would be true if the Lagrange multiplier
was introduced to the mean curvature: $H$ is fixed, but the equation for the Gauss curvature would get the right hand side including the Lagrange multiplier. This would leave $K$ unfixed.

One more obvious option to obtain equations (\ref{Gauss}), (\ref{Mean-r})
from variational principle is to add both $K$ and $H$ with independent Lagrange multipliers
to the action (\ref{S-WZ})
\begin{equation}\label{S-complete}\begin{array}
{c}\displaystyle
   S=\int
d^2\xi\sqrt{-h}\bigg[\,-\frac{1}{2\pi\alpha'}+\kappa K+\theta(\,2H+r^{-1})-\frac{\gamma}{3}(x,\mathfrak{n})\,\bigg]\,.
\end{array}\end{equation}
Here $\alpha', r, \gamma$ are constant parameters; $\kappa(\xi),\theta(\xi)$ --- Lagrange multipliers being dynamical fields on the world sheet.
Lagrangian equations read
\begin{equation}\label{EoM-theta-kappa}\begin{array}{rcl} \displaystyle
   \frac{\delta S}{\delta \kappa}&\equiv&\sqrt{-h}\,K=0\,;\\[3mm]\displaystyle
   \frac{\delta S}{\delta \theta}&\equiv& \sqrt{-h}\,(2H+r^{-1})=0\,;
 \end{array}\end{equation}
 \begin{equation}\label{EoM-x}\begin{array}{c}\displaystyle
   \frac{\delta S}{\delta x^{\,\mu}}\equiv
   -\sqrt{-h}\,h^{ab}\,[\,K\,\partial_a\,\kappa+(2H+r^{-1})\,\partial_a\,\theta\,]\partial_b\,x_\mu+\\[3mm]\displaystyle
   +\sqrt{-h}\,[\,\Delta\,\theta+\widetilde{\Delta}\,\kappa-2(K+Hr^{-1})\,\theta+(\pi\alpha')^{-1}H-\gamma\,]\,\mathfrak{n}_\mu=0\,,
\end{array}\end{equation}
where Laplacians $\Delta, \, \widetilde{\Delta}$ are defined by relations (\ref{LAP-def}).

Equations (\ref{EoM-theta-kappa}), (\ref{EoM-x}) are not  independent as there are three Noether identities between them. The first two identities obviously follow from the fact the action (\ref{S-complete}) is diffeomorphism-invariant,
i.e. the infinitesimal gauge transformations
\begin{equation}\label{Diff-transf}
\delta_\zeta x^\mu = -\zeta^a\partial_ax^\mu\, ,
\qquad
\delta_\zeta \kappa =- \zeta^a\partial_a\kappa\, ,
\qquad
\delta_\zeta \theta =-\zeta^a\partial_a\theta\,
\end{equation}
leave the action unchanged. Here, $\zeta^a (\xi)$ are the gauge parameters, being arbitrary vector fields on the world sheet. Corresponding Noether identities read
\begin{equation}\label{Diff-N-Id}
\partial_a x^\mu\,\frac{\delta S}{\delta x^\mu}
+\partial_a\kappa\, \frac{\delta S}{\delta \kappa}
+\partial_a \theta\, \frac{\delta S}{\delta \theta}\equiv0\, ,\quad a=0,1\,.
\end{equation}
Taking these identities into account, we see that among the three components of the vector equation (\ref{EoM-x}), only one, which is the projection onto the direction of the normal to the world sheet, cannot be reduced to the equations (\ref{EoM-theta-kappa}). Given relations (\ref{EoM-theta-kappa}), the projection of equation (\ref{EoM-x}) on the normal to the world sheet reads
\begin{equation}\label{Eq-theta-kappa}
  \Delta\,\theta+\widetilde{\Delta}\,\kappa+r^{-2}\theta-(2\pi\alpha'r)^{-1}-\gamma= \,0 \, .
  \end{equation}
Given the identities (\ref{Diff-N-Id}), we see that Lagrangian equations reduce to three second order scalar equations: (\ref{Gauss}), (\ref{Mean-r}) (Cf. (\ref{EoM-theta-kappa})) and (\ref{Eq-theta-kappa}). The first two equations unambiguously fix the world sheet as a right circular cylinder, given the periodic conditions for $x(\sigma)$. The cylinders of the same radius are distinguished by the position and direction of the axis, and this is the finite-dimensional space of initial data. Therefore, from a geometric point of view, the equations (\ref{EoM-theta-kappa}) cannot bring any local degree of freedom into the string evolution. On the other hand, the system (\ref{EoM-theta-kappa}) includes two second order scalar equations on three fields $x^\mu(\xi)$. Diffeomorphism, being zero order gauge transformation of $2d$ of three fields $x^\mu(\xi)$, do not gauge out any local DoF.  If there was no fourth order differential identity between equations (\ref{EoM-theta-kappa}), they would describe local DoF's. As we see, there is a hint from physics to the existence of the differential identity between Gauss curvature and mean curvature of the $2d$ surface in $3d$ Minkowski space. In $3d$ Euclidean space, the differential identity between $K$ and $H$ was first found in the article \cite{Olver1} in a special non-holonomic  frame. The authors named this identity ``Codazzi syzygy''. Discussion of the consequences of this identity for the geometry of $2d$ surfaces in $3d$ Euclidean space can be found in \cite{Olver2}. Explicitly covariant form of this identity has been unknown yet, nor is known the analogue for Minkowski space. In the Appendix, we deduce the differential identity (\ref{KH-identity}) between $K$ and $H$ in explicitly covariant form both for Minkowski and Euclidean space.  Since Lagrangian equations (\ref{EoM-theta-kappa}) include $K$ and $H$, and nothing else, relation (\ref{KH-identity}) can be considered as the Noether identity for the Lagrangian (\ref{S-complete})
\begin{equation}\label{NI2}
\big(- \mathcal{D}\,\Delta
+h^{ab}\big(\nabla_a \mathcal{D}\big)\,\nabla_b
-4\mathcal{D}^2\big)\,\frac{\delta S}{\delta \kappa}
+\big(\mathcal{D}\,\widetilde{\Delta}
- \mathfrak{B}{}^{ab}\big(\nabla_a\mathcal{D}\big)\, \nabla_b \big)\,\frac{\delta S}{\delta \theta}\equiv0\,,
\end{equation}
Given the Noether identity above, we can immediately obtain the corresponding infinitesimal gauge symmetry transformation
\begin{equation}\label{GT2}
\tilde{\delta}_\eta x^\mu=0\,,\quad \tilde{\delta}_\eta \kappa=-\Delta (\eta\,\mathcal{D})
-h^{ab}\nabla_a(\eta\,\nabla_b \mathcal{D})
-4\mathcal{D}^2\,\eta\,,\quad
\tilde{\delta}_\eta \theta=
\widetilde{\Delta}(\eta\,\mathcal{D})
+ \mathfrak{B}{}^{ab} \nabla_a( \eta\,\nabla_b\mathcal{D})\,.
\end{equation}
The gauge parameter $\eta$ is $2d$ and $3d$ scalar.  One can verify by direct variation  that the action (\ref{S-complete})  is indeed invariant under transformation (\ref{GT2}), with account of the identity (\ref{KH-identity}). The  gauge symmetry transformation (\ref{GT2}) does not affect $x^\mu$, so it cannot be a combination of diffeomorphism transformations (\ref{Diff-transf}).

Transformations (\ref{Diff-transf}), (\ref{GT2}) have the following commutation relations:
\begin{equation}\label{ALGEBRA}\begin{array}{c}\displaystyle
 [\delta_{\zeta_1}, \delta_{\zeta_2}]u^I=\delta_{\zeta_3}u^I\,,\quad \zeta_3^a=\zeta_1^b
 \partial_b \zeta_2^a-\zeta_2^b\partial_b \zeta_1^a\,,\\[3mm]\displaystyle
 [\tilde{\delta}_{\eta}, \delta_{\zeta}]\kappa=\tilde{\delta}_{-\partial_c\eta\zeta^c}\,\kappa\,,\quad
[\tilde{\delta}_\eta, \delta_\zeta]\theta=\tilde{\delta}_{-\partial_c\eta\zeta^c}\,\theta\,,
\\[3mm]\displaystyle
[\tilde{\delta}_\eta, \delta_\zeta]x^\mu=0\,,\quad  [\tilde{\delta}_{\eta_1}, \tilde{\delta}_{\eta_2}]u^I=0\,,
\end{array}\end{equation}
where
\begin{equation}\label{u-def}
   u^I=(x^\mu, \theta, \kappa)\,.
\end{equation}
So, the gauge algebra is the extension of the Lie algebra of vector fields by abelian ideal of the second order scalar transformations (\ref{GT2}).

Let us summarise the conclusions about the structure of Lagrangian equations (\ref{EoM-theta-kappa}), (\ref{EoM-x}) for the action (\ref{S-complete}).
Note that system of five Lagrangian equations (\ref{EoM-theta-kappa}), (\ref{EoM-x}) reduce to three scalar equations,
\begin{equation}\label{KH-eq}
  K=0\, ,\quad 2H+ r^{-1}=0,
\end{equation}
\begin{equation}\label{theta-kappa-eq}
  \Delta\,\theta+\widetilde{\Delta}\,\kappa+r^{-2}\theta-(2\pi\alpha'r)^{-1}-\gamma= \,0 \, .
\end{equation}
The equations (\ref{KH-eq}), (\ref{theta-kappa-eq}) are involutive as they admit no lower order consequences. Their gauge symmetries and identities are known, see (\ref{Diff-transf}), (\ref{GT2}), (\ref{KH-identity}). This allows us to count degree of freedom number $N_{DoF}$ in a manifestly covariant way making use of the formula (8) of article \cite{Involution2012},
\begin{equation}\label{DoFcount}
  N_{DoF}=\sum_{n=0}n(t_n-r_n-l_n) \, .
\end{equation}
Here, $t_n$ is the number of equations of $n$-th order in the system;  $r_n$ is the number of gauge symmetry transformations with $n$-th order derivatives of gauge parameter; $l_n$ is the number of gauge identities of the order $n$. The system (\ref{KH-eq}), (\ref{theta-kappa-eq}) includes three second order equations, $t_2=3$.  There are two zero order gauge transformations (\ref{Diff-transf}), $r_0=2$, and one second order gauge transformation (\ref{GT2}), $r_2=1$. There is one fourth order identity\footnote{The order of identity between equations is defined as the maximum of the sum of the order of the gauge generator and the order of the equation on which it acts \cite{Involution2012}. For example, consider Maxwell equations. Divergency of the left hand sides vanishes identically for these second order equations. The operator of divergency is of the first order, so it is the third order identity.} (\ref{KH-identity}), $l_4=1$. Substituting these specific numbers $t_n,\, r_n, \, l_n$ into DoF count formula (\ref{DoFcount})
\begin{equation}\label{DoFcount-Model}
  N_{DoF}= 2\cdot 3 - 2 \cdot 0 -1\cdot 2 - 4\cdot 1 = 0\, ,
\end{equation}
we see that the system is a topological theory having no local degree of freedom.
The same conclusion is confirmed by the Hamiltonian constrained analysis in the next Section.

Since the equations (\ref{KH-eq}), (\ref{theta-kappa-eq}) describe the topological system having no local degree of freedom,  let us discuss the global DoF's.  Given the  $(-+)$ signature of induced metric, and the periodic conditions for $x(\sigma)$, the equations (\ref{KH-eq}) define the geometry of the world sheet as the right circular cylinder with time-like axis. Canonical parameterization of the cylinder reads
\begin{equation}\label{param-cylinder}
  x^0= \tau \,, \quad x^1 = y^1+r\cos{\sigma}\,, \quad x^2 = y^2+r\sin{\sigma}\,.
\end{equation}
This solution corresponds to the following direction vector $n^\mu$ (\ref{PJ_ny_2}), (\ref{WS_ny}) of the cylinder axis
\begin{equation}\label{}
 n^0=\pm 1\,,\quad n^1=0\,, \quad n^2=0\,,
\end{equation}
and the constant space-like vector $y=(0,y^1,y^2), \, (n,y)=0$ defines the position of the axis.

Any other solution of equations (\ref{KH-eq}) is derived from this one by the change of local coordinates  $\tau,\sigma$ on the world sheet and by Poincar\'e transformation of the target space coordinates $x^\mu$.
As relations (\ref{param-cylinder}) define the world sheet modulo Poincar\'e transformations and world sheet reparametrization, we can substitute this solution for $x$ into the equation (\ref{theta-kappa-eq})  without loss of generality. As a result, we arrive at the linear equation for the Lagrange  multipliers $\kappa(\xi),\theta(\xi)$
\begin{equation}\label{eq-kappa-theta-linear}
  \ddot\kappa- r\ddot\theta+r^{-1}\acute{\,\theta}\acute{\phantom{\theta}\!\!\!}+ r^{-1}\theta-(2\pi\alpha')^{-1}-r\gamma=0\, .
\end{equation}
Upon substitution of solution  for $x$ (\ref{param-cylinder}) into the gauge transformation (\ref{GT2}) for $\kappa,\theta$ we get the linear second order transformation with constant coefficients,
\begin{equation}\label{GT2-linear}
  \tilde{\delta}_\omega\kappa= \frac{1}{4r^2}(\ddot\omega-r^{-2}\acute{\,\omega}\acute{\!}-r^{-2} \omega) \,, \quad \tilde{\delta}_\omega\theta=\frac{1}{4r^3}\ddot\omega \,.
\end{equation}
From the DoF count (\ref{DoFcount}), (\ref{DoFcount-Model}) we already know that the equation (\ref{eq-kappa-theta-linear}) does not describe any local degree of freedom. Now, let us discuss the general solution of (\ref{eq-kappa-theta-linear}) from the perspective  of global degree of freedom. Integrating equation (\ref{eq-kappa-theta-linear}) twice by $\tau$ we obtain the general solution for $\kappa (\tau,\sigma)$
in terms of arbitrary $\theta(\tau,\sigma)$
\begin{equation}\label{kappa-solution}
  \kappa=r\theta-r^{-1}\int\limits_0^{\tau}dt\int\limits_0^{t}d\tilde{t}\big(\acute{\,\theta}\acute{\phantom{\theta}\!\!\!}(\tilde{t},\sigma)+\theta(\tilde{t},\sigma)\big)+\frac{1}{2}\big(\gamma+(2\pi\alpha'r)^{-1}\big)\tau^2+F_0(\sigma)+F_1(\sigma)\tau\,,
\end{equation}
where $F_0, F_1$ are arbitrary periodic functions of $\sigma$. The functions $F_0, F_1$ do not describe any independent initial data because they can be removed by gauge transformation (\ref{GT2-linear}) with special gauge parameter
\begin{equation}\label{RemoveF}
  \omega=f_0(\sigma)+f_1(\sigma)\tau \, ,
\end{equation}
where $f_0,f_1$ obey equations
\begin{equation}\label{Eq-for-f}
  \acute{\,f}\acute{\phantom{f}\!\!\!\!}_0+f_0=F_0\, ,\quad \acute{\,f}\acute{\phantom{f}\!\!\!\!}_1+f_1=F_1\,.
\end{equation}
These equations have solution for any smooth $F_0,F_1$.
The only subtlety is that the gauge parameter (\ref{RemoveF}), with $f_0, f_1$ being solutions of equations (\ref{Eq-for-f}) with periodic right hand side, is not necessarily periodic function of $\sigma$. It can be quasi-periodic for the special $F=A\sin(\sigma + \varphi)$, where $A\,, \varphi$ are constants,
\begin{equation}\label{Eq-resonance}
 \acute{\,f}\acute{\phantom{f}\!\!\!\!}+f= A\sin(\sigma + \varphi)\, ,\quad\Rightarrow\quad  f= C\sin(\sigma+\psi) - \frac{A}{2}\sigma\cos(\sigma +\phi ), \quad \forall\,C,\psi - \text{const} .
\end{equation}
If we do not admit quasi-periodic gauge parameters resulting in periodic transformations for $\kappa$, then the constants $A, \phi$ cannot be removed from $\kappa$ by gauge transformation, and hence they would be global DoF's. We do not exclude quasi-periodic gauge parameters if they result in periodic transformations, so the general solution (\ref{kappa-solution}) for  $\kappa$ is a pure gauge.

\section{Hamiltonian formalism}
\label{sec:3}
In this section we develop constrained Hamiltonian formulation for the string model (\ref{S-complete}).

The action of the model (\ref{S-complete}) includes the second time derivatives of the world-sheet coordinates $x^\mu$, so the usual Dirac's method \cite{Dirac} cannot be immediately applied. We follow the general procedure of reference \cite{HDHF} that allows one to construct the constrained Hamiltonian formalism for a degenerate Lagrangian with higher derivatives.  At first, we introduce the extra variables  $\phi^\mu, \lambda_\kappa, \lambda_\theta,\, \nu^\mu$ to absorb the first time derivatives of $x^\mu$, $\kappa,\, \theta$, and the second derivatives of $x^\mu$,
\begin{equation}\label{dotx}
\dot{x}^\mu-\phi^\mu=0,\qquad
\dot{\phi}^\mu-\nu^\mu=0,
\end{equation}
\begin{equation}\label{dotkappa}
\dot{\kappa}-\lambda_\kappa=0,\qquad
\dot{\theta}-\lambda_\theta=0.
\end{equation}
The variables $x^\mu, \phi^\mu, \theta, \kappa$ are considered as the configurations space coordinates in constrained Hamiltonian formalism, while $\nu^\mu, \lambda_\kappa, \lambda_\theta$ play the role the Lagrange multipliers at the primary constraints.

Let us re-write the relevant geometric quantities in terms of these new variables, and introduce convenient abbreviations.
The determinant of the metric and normal read
\begin{equation}
h = (\phi,\phi)(\acute{x},\acute{x})-(\phi,\acute{x})(\phi,\acute{x}),
\end{equation}
\begin{equation}
\mathfrak{n}_\mu =
-\varepsilon_{\mu\nu\rho}
(-h)^{-\frac{1}{2}}
\phi^\nu \acute{x}^\rho.
\end{equation}
The components of the first fundamental form density, being expressed in terms of these new variables, are denoted as
$\mathcal{A}_\theta, \mathcal{B}_\theta, \mathcal{C}_\theta$,
\begin{equation}\label{Atheta}
h_{11} =(\acute{x},\acute{x})=
(-h)^{\frac{1}{2}}\mathcal{A}_\theta,
\end{equation}
\begin{equation}
h_{01} =(\phi,\acute{x})=
(-h)^{\frac{1}{2}}\mathcal{B}_\theta,
\end{equation}
\begin{equation}
h_{00} =(\phi,\phi)=
(-h)^{\frac{1}{2}}\mathcal{C}_\theta.
\end{equation}
For the components of the second fundamental form, being expressed in terms of these new variables, we introduce abbreviations
$\mathcal{A}_\kappa, \mathcal{B}_\kappa$,
\begin{equation}\label{Akappa}
\mathfrak{b}_{11}=(\acute{\,x}\acute{\!}\,,\mathfrak{n})=
(-h)^{\frac{1}{2}}\mathcal{A}_\kappa,
\qquad
\end{equation}
\begin{equation}
\mathfrak{b}_{01}=(\acute{\phi}\,,\mathfrak{n})=
(-h)^{\frac{1}{2}}\mathcal{B}_\kappa,
\qquad
\end{equation}
\begin{equation}
\mathfrak{b}_{00}=(\nu\,,\mathfrak{n}).
\qquad
\end{equation}
In the new variables, Christoffel symbols read
\begin{equation}
\Gamma_{01}^0
=h^{-1}[(\acute{\phi}\,,\phi)(\acute{x}\,,\acute{x})
-(\acute{\phi}\,,\acute{x})(\phi\,,\acute{x})],
\end{equation}
\begin{equation}
\Gamma_{01}^1
=h^{-1}[-(\acute{\phi}\,,\phi)(\phi\,,\acute{x})
+(\acute{\phi}\,,\acute{x})(\phi,\phi)],
\end{equation}
\begin{equation}
\Gamma_{11}^0
=h^{-1}[(\acute{\,x}\acute{\!}\,,\phi)(\acute{x}\,,\acute{x})
-(\acute{\,x}\acute{\!}\,,\acute{x})(\phi\,,\acute{x})],
\end{equation}
\begin{equation}
\Gamma_{11}^1
=h^{-1}[-(\acute{\,x}\acute{\!}\,,\phi)(\phi,\acute{x})
+(\acute{\,x}\acute{\!}\,,\acute{x})(\phi,\phi)].
\end{equation}
Upon exclusion of all the time derivatives in terms of these new variables, the Lagrangian (\ref{S-complete}) reads
$$
\mathcal{L}(x,\phi,\nu,\kappa,\theta)=
-(-h)^{\frac{1}{2}}\bigg(
\frac{1}{2\pi\alpha'}	
+\frac{\gamma}{3}(x,\mathfrak{n})\bigg)
$$
$$
+\kappa\, (-h)^{-\frac{1}{2}}
\bigg(
-(\nu\,,\mathfrak{n})(\acute{\,x}\acute{\!}\,,\mathfrak{n})
+(\acute{\phi}\,,\mathfrak{n})
(\acute{\phi}\,,\mathfrak{n})
\bigg)
$$
\begin{equation}\label{L1}
+\theta \bigg(
(-h)^{-\frac{1}{2}}
\big[
-(\nu\,,\mathfrak{n})(\acute{x}\,,\acute{x})
+2(\acute{\phi}\,,\mathfrak{n})(\phi\,,\acute{x})
-(\acute{\,x}\acute{\!}\,,\mathfrak{n})(\phi,\phi)
\big]
+(-h)^{\frac{1}{2}}r^{-1}
\bigg).
\end{equation}
To get the Hamiltonian action equivalent to the original second order theory (\ref{S-complete}), we add to the above Lagrangian conditions (\ref{dotx}), (\ref{dotkappa}) with corresponding Lagrange multiplies denoted $p_\mu, p_{\phi\mu}, p_\kappa, p_\theta$,
\begin{equation}\label{S-H}
  S_H=\int d^2\xi\big(p_\mu(\dot{x}{}^\mu-\phi^\mu)+ p_{\phi\mu}(\dot{\phi}{}^\mu-\nu^\mu)+p_\kappa(\dot{\kappa}-\lambda_\kappa)+p_\theta(\dot{\theta}-\lambda_\theta) +  \mathcal{L}(x,\phi,\nu,\kappa,\theta) \big).
\end{equation}
After substituting Lagrangian (\ref{L1}) into the above expression, and grouping the similar terms, we arrive at the action of the Hamiltonian system with primary constraints,
\begin{equation}\label{S-H1}
S_H=\int d^2\xi\big(p_\mu\dot{x}{}^\mu+ p_{\phi\mu}\dot{\phi}{}^\mu+p_\kappa\dot{\kappa}+p_\theta\dot{\theta} \, - \, \mathcal{H}_0(x,p,\phi,\kappa,\theta) - \nu^\mu \Phi^{(1)}_\mu  - \lambda_\kappa\Phi^{(1)}_\kappa
- \lambda_\theta\Phi^{(1)}_\theta\big) \, ,
\end{equation}
where primary Hamiltonian and constraints read
\begin{equation}\label{H0}
\mathcal{H}_0=(p,\phi)
+(-h)^{\frac{1}{2}}
\big[
-\kappa\,\mathcal{B}_\kappa^2
-\theta\,
(2\mathcal{B}_\kappa\mathcal{B}_\theta
-\mathcal{A}_\kappa\mathcal{C}_\theta
+ r^{-1})
+\frac{1}{2\pi\alpha'}
+\frac{\gamma}{3}(x,\mathfrak{n})
\big],
\end{equation}
\begin{equation}\label{inprc1}
\Phi^{(1)}_\mu =
p_\phi{}_\mu+\kappa \mathcal{A}_\kappa
\mathfrak{n}_\mu
+\theta \mathcal{A}_\theta\mathfrak{n}_\mu=0,
\end{equation}
\begin{equation}\label{inprc2}
\Phi^{(1)}_\kappa =p_\kappa = 0,
\end{equation}
\begin{equation}\label{inprc3}
\Phi^{(1)}_\theta =p_\theta = 0.
\end{equation}
Given the Hamiltonian action and primary constraints, we can apply Dirac-Bergmann algorithm to deduce secondary constraints and determine some of Lagrange multipliers.  Before doing that, it is convenient to regroup the primary constraints. The idea is to project $3d$ vectorial constraints onto normal and tangential directions to the world sheet, and to combine the momenta $p_\kappa, p_\theta$ with the elements of corresponding first or second fundamental forms defined by relations (\ref{Atheta}), (\ref{Akappa}) in order to facilitate separation of the first class constrainnts from the second class ones. The new basis of primary constraints read
\begin{equation}\label{constt10}
T^{(1)}_0 =(p_\phi,\phi),
\end{equation}
\begin{equation}\label{constt11}
T^{(1)}_1 =(p_\phi,\acute{x}),
\end{equation}
\begin{equation}\label{constt1minus}
T^{(1)}
= \mathcal{A}_\theta p_\kappa-\mathcal{A}_\kappa p_\theta,
\end{equation}
\begin{equation}\label{psi1n}
\Psi^{(1)}_\mathfrak{n}
=(p_\phi,\mathfrak{n}) +\kappa\mathcal{A}_\kappa+\theta\mathcal{A}_\theta,
\end{equation}
\begin{equation}\label{psi1plus}
\Psi^{(1)}
= \mathcal{A}_\kappa p_\kappa+\mathcal{A}_\theta p_\theta .
\end{equation}
Non-vanishing Poisson brackets of these constraints read
\begin{equation}\label{t10t11}
\{T^{(1)}_0,T^{(1)}_1\}=T^{(1)}_1,
\end{equation}
\begin{equation}\label{t10t1}
\{T^{(1)}_0,T^{(1)}\}=T^{(1)},
\end{equation}
\begin{equation}
\{T^{(1)}_0,\Psi^{(1)}_\mathfrak{n}\}
=\Psi^{(1)}_\mathfrak{n},
\end{equation}
\begin{equation}
\{T^{(1)}_0,\Psi^{(1)}\}=\Psi^{(1)},
\end{equation}
\begin{equation}
\{\Psi^{(1)}_\mathfrak{n},T^{(1)}\}
=(-h)^{-\frac{1}{2}}
\Gamma^0_{11}
\Upsilon
(\mathcal{A}_\kappa T^{(1)}
-\mathcal{A}_\theta\Psi^{(1)}),
\end{equation}
\begin{equation}\label{afc26}
\{\Psi^{(1)}_\mathfrak{n},\Psi^{(1)}\}
=
(-h)^{-\frac{1}{2}}
\Gamma^0_{11}
\Upsilon
(\mathcal{A}_\theta T^{(1)}
+\mathcal{A}_\kappa\Psi^{(1)})
+\Upsilon^{-1},
\end{equation}
where the abbreviation is used
\begin{equation}
\Upsilon^{-1}=\mathcal{A}_\kappa^2+\mathcal{A}_\theta^2 \, .
\end{equation}
This quantity is strictly positive because
$(\acute{x},\acute{x})=
(-h)^{\frac{1}{2}}\mathcal{A}_\theta>0$.

As one can see, primary constraints $\Psi^{(1)}_\mathfrak{n}$
(\ref{psi1n}), $\Psi^{(1)}$ (\ref{psi1plus}) are of the second class.
The remaining primary constraints
$T^{(1)}_0$ (\ref{constt10}), $T^{(1)}_1$ (\ref{constt11}),
$T^{(1)}$ (\ref{constt1minus}) Poisson commute to each other and with $\Psi^{(1)}_\mathfrak{n}, \Psi^{(1)}$.

Introduce the Hamiltonian $\mathcal{H}^{(1)}$,
\begin{equation}\label{H1mod}
\mathcal{H}^{(1)}=\mathcal{H}_0
+{\mu}_\mathfrak{n}\Psi^{(1)}_\mathfrak{n}
+{\mu}\,\Psi^{(1)}
+{\lambda}^0 T^{(1)}_0
+{\lambda}^1 T^{(1)}_1
+{\lambda}\, T^{(1)},
\end{equation}
where $\mathcal{H}_0$ is defined by (\ref{H0}),
\begin{equation}
H^{(1)}=\int d\sigma \mathcal{H}^{(1)}.
\end{equation}
Let us examine conservation of primary constraints,
\begin{equation}\label{ALG}\begin{array}{c}\displaystyle
\dot{\Psi}^{(1)}_\mathfrak{n}
=\{\Psi^{(1)}_\mathfrak{n}\,,H^{(1)}\}\approx0,\quad
\dot{\Psi}^{(1)}=\{\Psi^{(1)}\,,H^{(1)}\}\approx 0,
\\[3mm]\displaystyle
\dot{T}^{(1)}_0=\{T^{(1)}_0\,,H^{(1)}\}\approx 0,\quad
\dot{T}^{(1)}_1=\{T^{(1)}_1\,,H^{(1)}\}\approx 0,\quad
\dot{T}^{(1)}=\{T^{(1)}\,,H^{(1)}\}\approx 0 .
\end{array}\end{equation}
These conditions define two Lagrange multipliers at the second class constraints,
\begin{equation}
\overline{{\mu}_\mathfrak{n}}=
\Upsilon(-h)^{\frac{1}{2}}
[\mathcal{A}_\kappa\mathcal{B}_\kappa^2
+2\mathcal{A}_\theta\mathcal{B}_\kappa\mathcal{B}_\theta
-\mathcal{A}_\kappa\mathcal{A}_\theta\mathcal{C}_\theta
+\mathcal{A}_\theta r^{-1}],
\end{equation}
\begin{equation}
\overline{\mu}=\Upsilon
[(p,\mathfrak{n})
+\kappa\,\acute{\mathcal{B}}_\kappa
+\theta\,\acute{\mathcal{B}}_\theta
+2\,\acute{\kappa}\,\mathcal{B}_\kappa
+2\,\acute{\theta}\,\mathcal{B}_\theta
+\frac{\gamma}{3}\,(x\,,\acute{x}\,,\mathfrak{n})]\, ,
\end{equation}
and lead to three secondary constraints
\begin{equation}
\mathcal{H}_0\approx0,
\end{equation}
\begin{equation}
(p,\acute{x})
-(p_\phi,\acute{\phi})
-2(-h)^{\frac{1}{2}}\mathcal{B}_\kappa
[\kappa\mathcal{A}_\kappa+\theta\mathcal{A}_\theta]\approx0,
\end{equation}
\begin{equation}
(-h)^{\frac{1}{2}}
[-\mathcal{A}_\theta\mathcal{B}_\kappa^2
+2\mathcal{A}_\kappa\mathcal{B}_\kappa\mathcal{B}_\theta
-\mathcal{A}_\kappa^2\mathcal{C}_\theta
+\mathcal{A}_\kappa r^{-1}]\approx 0\, .
\end{equation}
The Hamiltonian $\mathcal{H}_0$ is a secondary constraint, as it should be, given the reparameterisation invariance of the action.
Lagrange multipliers  ${\lambda}^0, {\lambda}^1,
{\lambda}$
are not determined by conditions (\ref{ALG}).

The listed above three secondary constraints are not a pure first class. Combining them with the primary constraints, we get three secondary first class constraints,
\begin{equation}\label{constt20}
T^{(2)}_0 = \mathcal{H}_0
+\overline{\mu_\mathfrak{n}}\,\Psi^{(1)}_\mathfrak{n}
+\overline{\mu}\,\Psi^{(1)},
\end{equation}
\begin{equation}\label{constt21}
T^{(2)}_1
=(p,\acute{x})+(p_\phi,\acute{\phi})
+ p_\kappa\,\acute{\kappa}
+ p_\theta\,\acute{\theta},
\end{equation}
$$
T^{(2)}=(-h)^{\frac{1}{2}}
\big(
-\mathcal{A}_\theta\mathcal{B}_\kappa^2
+2\mathcal{A}_\kappa\mathcal{B}_\kappa\mathcal{B}_\theta
-\mathcal{A}_\kappa^2\mathcal{C}_\theta
+\mathcal{A}_\kappa r^{-1}
\big)
$$
\begin{equation}\label{constt2minus}
+\left(
\frac{\partial}{\partial \sigma}
-\Gamma^1_{11}
\right)
(
2\mathcal{B}_\theta p_\kappa
-2\mathcal{B}_\kappa p_\theta
)
-\Gamma^0_{11}
(\mathcal{C}_\theta p_\kappa
-\mathcal{C}_\kappa p_\theta).
\end{equation}
Above, we have used the abbreviation
\begin{equation}
\mathcal{C}_\kappa
=(-h)^{-\frac{1}{2}}\,\overline{\mu_\mathfrak{n}}
=\Upsilon
[\mathcal{A}_\kappa\mathcal{B}_\kappa^2
+2\mathcal{A}_\theta\mathcal{B}_\kappa\mathcal{B}_\theta
-\mathcal{A}_\kappa\mathcal{A}_\theta\mathcal{C}_\theta
+\mathcal{A}_\theta r^{-1}].
\end{equation}
Given three secondary constraints, we examine their conservation.
Constraints $T^{(2)}_0$,
$T^{(2)}_1$ conserve identically.\\
Conservation of $T^{(2)}$,
\begin{equation}
\dot{T}^{(2)}=\{T^{(2)}\,,H^{(1)}\}\approx 0,
\end{equation}
results in the tertiary constraint
$$
(-h)^{\frac{1}{2}}\bigg(
(2\mathcal{B}_\kappa\mathcal{C}_\theta
-2\mathcal{B}_\theta\mathcal{C}_\kappa)
(\acute{\mathcal{A}}_\kappa
+\Gamma^0_{01}\mathcal{A}_\kappa
-2\Gamma^0_{11}\mathcal{B}_\kappa
-\Gamma^1_{11}\mathcal{A}_\kappa)
$$
\begin{equation}
+(
2\mathcal{B}_\kappa\mathcal{B}_\theta
-2\mathcal{A}_\kappa\mathcal{C}_\theta
+ r^{-1}
)(\acute{\mathcal{B}}_\kappa
-\Gamma^1_{01}\mathcal{A}_\kappa
-\Gamma^0_{11}\mathcal{C}_\kappa)
\bigg)\approx 0.
\end{equation}
This is not a pure first class constraint. By adding appropriate combination of the primary and secondary constraints, we get the first class constraint
$$
T^{(3)}=
(-h)^{\frac{1}{2}}\bigg(
(2\mathcal{B}_\kappa\mathcal{C}_\theta
-2\mathcal{B}_\theta\mathcal{C}_\kappa)
(\acute{\mathcal{A}}_\kappa
+\Gamma^0_{01}\mathcal{A}_\kappa
-2\Gamma^0_{11}\mathcal{B}_\kappa
-\Gamma^1_{11}\mathcal{A}_\kappa)
$$
$$
+(
2\mathcal{B}_\kappa\mathcal{B}_\theta
-2\mathcal{A}_\kappa\mathcal{C}_\theta
+ r^{-1}
)(\acute{\mathcal{B}}_\kappa
-\Gamma^1_{01}\mathcal{A}_\kappa
-\Gamma^0_{11}\mathcal{C}_\kappa)
$$
\begin{equation}\nonumber
-\big(\Gamma^0_{11}
(\mathcal{A}_\kappa\mathcal{C}_\kappa
+\mathcal{A}_\theta\mathcal{C}_\theta )
\Upsilon
+\Gamma^1_{01}\big)
\big(-\mathcal{A}_\theta\mathcal{B}_\kappa^2
+2\mathcal{A}_\kappa\mathcal{B}_\kappa\mathcal{B}_\theta
-\mathcal{A}_\kappa^2\mathcal{C}_\theta
+\mathcal{A}_\kappa r^{-1}\big)
\bigg)
\end{equation}
\begin{equation}\label{constt3minus}
+(-h)^{\frac{1}{2}}r^{-2}\,p_\kappa
+\left(
\frac{\partial}{\partial \sigma}
-\Gamma^0_{01}
-\Gamma^1_{11}
\right){F}.
\end{equation}
Here, we use abbreviation
\begin{equation}
F
=
-\left(\frac{\partial}{\partial \sigma}
-\Gamma^0_{01}\right)
(\mathcal{C}_\theta p_\kappa
-\mathcal{C}_\kappa p_\theta)
+\Gamma^1_{01}(2\mathcal{B}_\theta p_\kappa
-2\mathcal{B}_\kappa p_\theta).
\end{equation}
Conservation of this constraint does not lead to further constraints, no does it define any Lagrange multiplier.
Finally we have 2 second class constraints $m_2=2$, and seven first class ones $m_1=7$. The phase space dimension is sixteen, $n=16$.
Hamiltonian local DoF count read
\begin{equation}\label{DoF-H}
  N_{DoF}=n-2m_1-m_2=16-2\cdot 7 - 2=0 \, ,
\end{equation}
in agreement with covariant DoF count (\ref{DoFcount}), (\ref{DoFcount-Model}).

As we have seen in this section, the constrained analysis confirms that the model is indeed a topological field theory. In principle, the Hamiltonian formalism should allow one to derive gauge symmetry from involution relations of  constraints. For this particular model, this is a cumbersome problem, because there are three generations of first class constraints, and there are higher derivatives of gauge parameters involved, and the second class constraints further complicate the issue.  The gauge transformations  (\ref{Diff-transf}), (\ref{GT2}) are not immediately reproduced in a canonical way by the specific basis of first class constraints
(\ref{constt10})-(\ref{constt1minus}),
(\ref{constt20})-(\ref{constt2minus}),
(\ref{constt3minus}).
To derive this symmetry from the Hamiltonian formalism one would have to find an appropriate generating set of constraints which not immediately seen. Another related important issue is quantisation of this string model. Since the constraints are known, the classical Hamiltonian BRST complex can be constructed in the usual way. To find the BRST cohomology in the space of quantum states one would have to construct an appropriate operator realisation of
the constraints. The classical constraints are not polynomials of the original canonical variables, and their commutation relations involve field dependent structure functions. This does not seem appropriate set of variables and constraints to solve for the equation $Q_{BRST}|\Psi>=0$.
One more open issue related with quantisation is the potential anomaly.  The gauge symmetry includes Virasoro algebra, so the central extension can be expected upon quantisation, while no reason is seen in advance why it should have critical value to  avoid anomaly.
At classical level, this string is a topological $2d$ field theory whose global modes describe dynamics of a single massive particle with spin.
We may expect that the space of physical states of the string can be eventually identified with the irreducible representation of the Poincar\'e group, but it seems not realistic to see that in terms of original variables and original set of constraints. So, this is the issue for further studies.

\section{Conclusion}
\label{sec:4}
Let us summarise the results, discuss open issues, and further perspectives.

In this article we propose the action (\ref{S-complete}) for the string in $3d$ Minkowski space. The least action principle leads to the equations (\ref{Gauss}), (\ref{Mean-r}) for the world sheet and the equation (\ref{Eq-theta-kappa}) for the Lagrange multipliers.  The latter equation enjoys the second order gauge symmetry (\ref{GT2}) which gauges out all the DoF's from the Lagrange multipliers. The equations (\ref{Gauss}), (\ref{Mean-r}) also describe the topological $2d$ field theory without local DoF.  The covariant local DoF count (\ref{DoFcount}), (\ref{DoFcount-Model}) is confirmed by the Hamiltonian constrained analysis.  The world sheet defined by of the string described by equations (\ref{Gauss}), (\ref{Mean-r}) is a right circular cylinder with the time-like axis and fixed radius.  The global DoF's of this world-sheet are related to the direction and position of the cylinder. These global  DoF's describe a single irreducible massive particle with spin, as we explain in the introduction.  The direction of cylinder axis is determined by particle's conserved total momentum, while the position of the axis is determined by conserved total angular momentum  (\ref{PJ_ny}), (\ref{WS_ny}).  In this way one can see that truncation of the string dynamics to a single particle at the level of action means inclusion of the two terms (\ref{S-complete}) which fix the mean curvature and Gauss curvature of the world sheet.

Note that the dynamics of a spinning particle is localized on a specific world sheet for any irreducible representation; this phenomenon is not a special feature of the massive case. For the continuous spin irreducible representation, for example, the particle world sheet is a parabolic cylinder 
If we define this world sheet by the field equations of corresponding string, it should be a topological $2d$ field theory whose global DoF's describe this particle. The equations, action, and gauge symmetry of such a string is unknown yet.

Even more mysterious is the problem of extending this theory to higher dimensions.  In $d=3$, the irreducibility of the massive Poncar\'e group representation leads to the conclusion that classical dynamics of spinning particle is localised at the world sheet (\ref{WS_PJ_2}) being the right circular cylinder with the time-like axis. The same logic as presented in the introduction leads to the conclusion that the dynamics of irreducible massive particle with spin is localised at toroidal cylinder $\mathbb{R}\times \mathbb{T}^D,\, D=[(d-1)/2] $ in $d$-dimensional Minkowski space \cite{WS2017}. If this world sheet is defined by the field equations that fix principal curvatures of the surface, similar to (\ref{Gauss}), (\ref{Mean-r}), then there should exist higher dimensional analogues of the differential identity (\ref{KH-identity}) between mean curvature and Gauss curvature of $2d$ surface in $3d$ Minkowski space. These identities are still unknown in $d>3$, to the best of our knowledge.
Thus, the idea of describing dynamics of classical spinning particles by world sheets suggests the existence of new theorems in the old field of differential surface geometry in (pseudo)Euclidean spaces. It further leads to extension of the $3d$ string gauge algebra (\ref{GT2}), (\ref{ALGEBRA}) by further higher order transformations. DoF evaluation suggests reducibility of this gauge algebra.

Another set of problems in the world sheet theory of spinning particles concerns the inclusion of interactions with external fields.
Interaction with constant electromagnetic field is included in the world sheet algebraic equations in the article \cite{WS2022_2},
though it remains unclear how this could affect equations of motion for the corresponding string (\ref{Gauss}), (\ref{Mean-r}). Inclusion of interaction of the string at the level of action (\ref{S-complete}) would mean to modification of the gauge symmetry (\ref{GT2}) in the way which is consistent with general metric $g_{\mu\nu}$ instead of Minkowski metric. The gauge symmetry is immediately connected with the identity  (\ref{KH-identity}) which should be modified to account for the curvature of the $3d$ space.
This modification is still unknown to the best of our knowledge.  So, the problem of inclusion consistent interactions of spinning particles with gravity transforms into the pure geometric problem of inclusion the curvature of $3d$ manifold into the equations (\ref{Gauss}), (\ref{Mean-r}) in the way such that they still obey the deformed identity (\ref{KH-identity}).


\vspace{0.2 cm}
\textbf{Acknowledgements} The work was supported by the Ministry of Science and Higher Education of the Russian
Federation (Project No. FSWM-2020-0033). The authors thank A.A.~Sharapov for valuable discussions. We benefited from comments by A.V.~Penskoi and Th.Th.~Voronov concerning the geometric issues related to this article.

\appendix
\section{}
\label{sec:app}
\noindent
In this Appendix we deduce the differential identity between Gauss curvature $K$ and mean curvature $H$ of $2d$ surface in $3d$ Minkowski, or Euclidean space.

\noindent
To uniformly consider Euclidean and pseudo-Euclidean spaces, we introduce the sign factor $\varepsilon=\pm 1.$  Minkowski space corresponds to $\varepsilon=1$, while $\varepsilon=-1$ for the Euclidean case.
In this notation the metric of dimensional (pseudo-)Euclidean space reads
$
\eta_{\alpha\beta}=\mbox{diag}(-\varepsilon, 1,1)
$.
To account for the signature, we modify the definition of the tensor
$\mathfrak{B}^{ab}$ (\ref{B}),
$$\mathfrak{B}^{ab}=\varepsilon\,
h^{-1}\varepsilon^{ac}\varepsilon^{bd}\mathfrak{b}_{cd}\, . $$
When  $\varepsilon=1$, this definition coincides with (\ref{B}).
Given $\mathfrak{B}^{ab}$, the Laplacians $\Delta,\,\widetilde{\Delta}$
are uniformly defined by relations (\ref{LAP-def}).

Let us explain the general strategy of deducing the differential identity between Gauss curvature $K$ and mean curvature $H$.
We know from the physical reasons of Section \ref{sec:2} that there should be fourth order differential identity between $K$ and  $H$.
This means that scalar linear combination of second derivatives of $K$ and $H$ should identically reduce to a combination of Gauss and mean curvature and their first derivatives.
Note that the general linear combination of second derivatives of $K$ and $H$ involve the second and first order derivatives of second fundamental form, and $\mathfrak{b}_{ab}$ itself.
The first step is to find the combination of second derivatives of $K$ and $H$ such that does not involve the second derivatives of $\mathfrak{b}_{ab}$.
This problem reduces to solving linear algebraic equations for the coefficients of the combination of second derivatives of the curvature. The solution for these coefficients is unique, modulo overall factor.  As the second step, we consider the scalar linear combination of $\nabla^2 K$ and $\nabla^2 H$ such that does not involve  $\nabla^2 \mathfrak{b}$.
We see that it reduces to a bilinear combination of the first derivatives of $\mathfrak{b}_{ab}$ denoted by $\mathcal{R}$,
and the polynomial of curvatures.
As the third step, we assume that $\mathcal{R}$ is a linear combination of the bilinear forms of the first derivatives of $K$ and $H$.
This again leads to a system of algebraic linear equations for the expansion coefficients  of $\mathcal{R}$ into bilinear combinations of the first derivatives of the curvatures.
The solution exists, and it is unique.
This finally demonstrates that certain linear combination of second  derivatives of $K$ and $H$ identically reduces to the first derivatives of the curvatures and the function of them.

Now let us begin to implement the above plan.
Given the Gauss and Peterson-Mainardi-Codazzi identities (\ref{GPMCe}),
one can see that the combination $2\widetilde{\Delta} H -\Delta K$
does not involve the second order derivatives of $\mathfrak{b}_{ab}$.
It includes at most the first derivatives of the second fundamental form,
\begin{equation}\label{t1}
2\widetilde{\Delta} H
-\Delta K
=
\mathcal{R}
+4 \mathcal{D} K ,
\end{equation}
where $\mathcal{D}$ is defined by the formula (\ref{D-def}),
$\mathcal{R}$ is bilinear in the first derivatives of $\mathfrak{b}$,
\begin{equation}\label{R}
  \qquad \mathcal{R}=-h^{ab}\nabla_a\mathfrak{B}{}^{cd}\,
\nabla_b\mathfrak{b}_{cd}\,.
\end{equation}
Let us assume that $\mathcal{R}$ (\ref{R}) reduces to bilinear combinations of the firs derivatives of $K$ and $H$,
\begin{equation}\label{Rcoeff}
\mathcal{R}=
A^{ab}\nabla_a H\nabla_bH
+
B^{ab}\nabla_aK\nabla_bK
+
C^{ab}\nabla_aH\nabla_bK,
\end{equation}
where
$A^{ab}$, $B^{ab}$, $C^{ab}$ -- are unknown coefficients. The linear space of derivatives $\nabla_a H, \nabla_b K$ is four-dimensional, so the linear space of quadratic forms is ten-dimensional.
Hence, there 10 independent unknown coefficients $A^{ab}$, $B^{ab}$, $C^{ab}$ in relation (\ref{Rcoeff}).
There are four independent components of the first derivatives of $\nabla_a\mathfrak{b}_{bc}$,
given the Peterson-Mainardi-Codazzi identity (\ref{GPMCe}). In this way one can see that assumption (\ref{Rcoeff}) about the structure of $\mathcal{R}$ (\ref{R}) leads to the system of ten linear equations for ten unknown coefficients $A^{ab}$, $B^{ab}$, $C^{ab}$. The solution is unique, and it reads
\begin{equation}\label{Rres}
\mathcal{R}=\frac{\nabla_a\mathcal{D}}{\mathcal{D}}
\left(2\mathfrak{B}^{ab}\nabla_bH-h^{ab}\nabla_bK\right).
\end{equation}
Substituting (\ref{Rres}) in (\ref{t1}) we arrive at the identity
\begin{equation}\label{KH-identity}
\begin{gathered}
2\widetilde{\Delta} H-\Delta K-
\frac{\nabla_a\mathcal{D}}{\mathcal{D}}
\left(2\mathfrak{B}^{ab}\nabla_bH-h^{ab}\nabla_bK\right)
-4\mathcal{D}K=0\, .
\end{gathered}
\end{equation}
Introducing the principal curvatures,
 \begin{equation}\label{k1k2}
   2H=k_1+k_2\, ,\quad K=k_1\cdot k_2\, ,
 \end{equation}
 we can reformulate the identity (\ref{KH-identity}) in terms of $k_1,k_2$,
\begin{equation}\label{kk-identity}
\begin{gathered}
(\mathfrak{B}{}^{ab}-h^{ab}k_2)\nabla_a\nabla_bk_1
+(\mathfrak{B}{}^{ab}-h^{ab}k_1)\nabla_a\nabla_bk_2=
\\[3mm]
=\frac{2(\mathfrak{B}{}^{ab}-h^{ab}k_2)\nabla_ak_1\nabla_bk_1
-2(\mathfrak{B}{}^{ab}-h^{ab}k_1)\nabla_ak_2\nabla_bk_2}{(k_1-k_2)}
+(k_1-k_2)^2k_1k_2.
\end{gathered}
\end{equation}

\end{document}